\documentclass[runningheads]{llncs}
\usepackage[T1]{fontenc}
\usepackage{graphicx}
\usepackage[colorlinks]{hyperref}

\hypersetup{
     colorlinks=true,
     linkcolor=red,
     filecolor=blue,
     citecolor =blue,      
     urlcolor=cyan,
     }
\usepackage{wrapfig,lipsum}
\usepackage{amsmath,bm}
\usepackage{url}            
\usepackage{booktabs}       
\usepackage{amssymb}
\usepackage{subfig}
\usepackage{cite}  
\usepackage{xcolor}
\usepackage{float}
\usepackage{amsmath}
\usepackage{multirow}
\usepackage{makecell}
\usepackage{color}
\usepackage{mathtools}
\usepackage{booktabs}
\usepackage{comment}
\usepackage{ifthen}
\usepackage{colortbl}
\usepackage{float} 
\usepackage{pifont}
\usepackage{enumitem}
\usepackage{threeparttable}
\usepackage{booktabs}
\usepackage{relsize}
\usepackage[T1]{fontenc}
\usepackage{wrapfig}
\usepackage{subfig}

\usepackage{orcidlink}
\usepackage[misc]{ifsym}

\definecolor{thalamus}{HTML}{67cc5c} % Medium Sea Green
\definecolor{cavum}{HTML}{cde01d} % Inch Worm
\definecolor{fossa}{HTML}{97d83e} % Yellow Green
\definecolor{head}{HTML}{fde724} % Saffron
\usepackage{xspace}

\def\dime{$\text{DiME}$\xspace} 
\def\difficeone{$\text{Diff-ICE}_{1}$\xspace} 
\def\difficeonenoisy{$\text{Diff-ICE}_{1}$-$\text{x}_{t}$\xspace}
\def\diffice{$\text{Diff-ICE}$\xspace}

\begin{document}
\title{Diffusion-based Iterative Counterfactual Explanations for Fetal Ultrasound \\ Image Quality Assessment}
\titlerunning{Iterative Counterfactual Explanations for Fetal Ultrasound Image Quality}

\author{Paraskevas Pegios\inst{1,4}\textsuperscript{(\Letter)}\orcidlink{0009-0005-1471-4850} 
\and
Manxi Lin\inst{1}\orcidlink{0000-0003-3399-8682}
\and
Nina Weng\inst{1}
\and
Morten Bo Søndergaard Svendsen\inst{2}\orcidlink{0000-0002-4492-3750}
\and
Zahra Bashir\inst{3}\orcidlink{0000-0002-2497-282X}
\and
Siavash Bigdeli\inst{1}
\and
Anders Nymark Christensen\inst{1}\orcidlink{0000-0002-3668-3128}
\and
Martin Tolsgaard\inst{2}\orcidlink{0000-0001-9197-5564}
\and
Aasa Feragen\inst{1,4}\orcidlink{0000-0002-9945-981X}
}
\authorrunning{P. Pegios et al.}

\institute{Technical University of Denmark, Kongens Lyngby, Denmark\\ \email{\{ppar,afhar\}@dtu.dk} \and
Region Hovedstaden Hospital, Copenhagen, Denmark\\ \and Slagelse Hospital, Copenhagen, Denmark\\ \and
Pioneer Centre for AI, Copenhagen, Denmark}

\maketitle           
\begin{abstract}
Obstetric ultrasound image quality is crucial for accurate diagnosis and monitoring of fetal health. However, acquiring high-quality standard planes is difficult, influenced by the sonographer's expertise and factors like the maternal BMI or fetus dynamics. In this work, we explore diffusion-based counterfactual explainable AI to generate realistic, high-quality standard planes from low-quality non-standard ones. Through quantitative and qualitative evaluation, we demonstrate the effectiveness of our approach in generating plausible counterfactuals of increased quality. This shows future promise for enhancing training of clinicians by providing visual feedback and potentially improving standard plane quality and acquisition for downstream diagnosis and monitoring. 
\end{abstract}
\section{Introduction}

\begin{figure}[b]
	\centering
	\subfloat{\includegraphics[width=\linewidth]{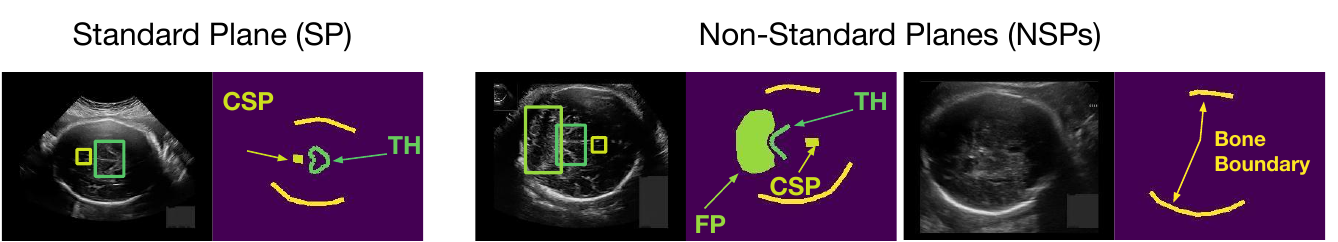}}

	\caption{Standard planes (SPs) are defined as particular anatomical planes through the body; here we show examples of high-quality SPs (left) and low-quality non-standard planes (NSPs) (right) for the fetal head. A fetal head SP should show \textcolor{thalamus}{thalamus (TH)},  \textcolor{cavum}{cavum septi pellucidi (CSP)}, but \emph{not} \textcolor{fossa}{fossa posterior (FP)}. The bone boundaries should support the correct placement of calipers.}
 \label{fig:sp}
\end{figure}

The quality of obstetric ultrasound screening images is crucial for the clinical downstream tasks, involving fetal growth estimation~\cite{wu2017fuiqa}, preterm birth prediction~\cite{pegios2023leveraging}, as well as abnormality detection~\cite{olsen2024unsupervised}.
If the captured anatomical planes are not precise or the anatomical regions to be measured are not visualized well, then the measurements and estimated outcomes will be incorrect (Fig.~\ref{fig:sp}). To standardize image quality, international ultrasound societies establish precise criteria for defining fetal ultrasound quality~\cite{salomon2019isuog}.  However, the acquisition of high-quality images is hampered by both the clinician's level of training and physical characteristics such as maternal BMI or the fetus position. As a result, acquiring \emph{standard planes} of sufficiently high quality can be very challenging.

As a step towards solving this problem, we investigate \emph{counterfactual visual explanations} for fetal ultrasound quality assessment. Given a low-quality non-standard plane (NSP) as input, we estimate a higher quality standard plane (SP) of the same anatomy.  While rooted in explainable AI, our approach has potential applications beyond model explainability by enabling impactful generative AI applications %~\cite{mendez2023approaches} 
such as synthetic data generation~\cite{wang2024generative}, personalized clinical training~\cite{men2023towards}, and computer-assisted plane acquisition~\cite{men2025scanahead}, with the ultimate goal of supporting reliable diagnosis in cases where SPs are difficult to obtain.

Our method builds on state-of-the-art diffusion-based counterfactual explanation approaches~\cite{jeanneret2022diffusion,jeanneret2023adversarial,weng2024fast,sobieski2024rethinking}. However, these are typically designed to make minimal, highly localized changes. This assumption does not align well with fetal ultrasound quality assessment, where improving quality often requires broader modifications, as correcting a NSP may require more global adjustments~\cite{salomon2019isuog}. To this end, we contribute 1) an iterative diffusion-based counterfactual generation method that drives toward higher confidence counterfactuals, achieving broad changes to the input, and 2) an extensive evaluation that demonstrates that we can generate plausible, high-quality counterfactual paths for fetal ultrasound.

\section{Related Work}
Deep learning methods have been widely used for ultrasound image quality assessment\cite{wu2017fuiqa,zhao2023memory,lin2024learning}. Generative models have further supported this task by improving fine-grained classification~\cite{maack2022gans} and creating training material for sonographers~\cite{lee2020generating,men2023towards}. Very recently, ScanAhead~\cite{men2025scanahead}, a GAN-based method with domain adaptation, was proposed to predict high-quality standard planes from scanning videos, incorporating 3D probe movement information. The success of denoising diffusion probabilistic models (DDPMs)~\cite{ho2020denoising,dhariwal2021diffusion} enabled the creation of highly realistic fetal ultrasound images~\cite{iskandar2023towards,wang2024generative} and enhanced out-of-distribution detection~\cite{mishra2023dual,olsen2024unsupervised}. In this paper, we rely solely on image data and existing ultrasound databases to explore a guided diffusion~\cite{dhariwal2021diffusion} counterfactual iterative framework that aims to predict the path from low-quality to high-quality images.

Counterfactual explanations try to answer the question: How does an image look if it is classified differently by a given classifier? Different from adversarial examples, counterfactuals should be realistic, i.e., close to the
data manifold, which is usually approximated with generative models~\cite{pegios2024counterfactual}. In computer vision, different methods have been proposed, including %VAE-based~\cite{cohen2021gifsplanation}, 
GAN-based~\cite{singla2023explaining} 
and diffusion-based~\cite{sanchez2022healthy,jeanneret2022diffusion,jeanneret2023adversarial,sobieski2024rethinking,weng2024fast} approaches. Yet, these are applied to tasks where changes are highly localized. In our work, we leverage recent advances in diffusion guidance~\cite{he2023manifold,weng2024fast} to apply a computationally feasible iterative approach, rather than sparsity~\cite{jeanneret2022diffusion}, inpainting~\cite{jeanneret2023adversarial,weng2024fast}, or region constraint~\cite {sobieski2024rethinking} to refine diffusion-based counterfactuals. This allows us to achieve more global counterfactual changes.

\section{Method}

\begin{figure}[t]
	\centering
  \subfloat{\includegraphics[scale=0.555]{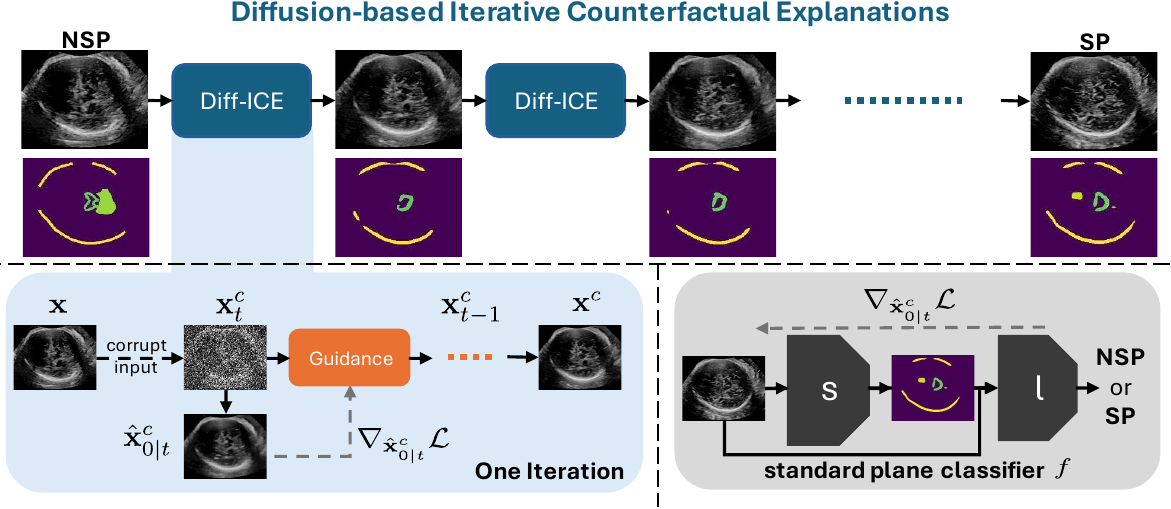}}

\caption{\textbf{Top:} Each iteration uses previous output to enhance counterfactual confidence. \textbf{Bottom-left:} Efficient gradient estimation at each time step $t$ for one iteration. \textbf{Bottom-right:} Standard plane classifier and guiding gradient flow.}
 
 \label{fig:method}
\end{figure}

\subsection{Preliminaries on Diffusion Models}
DDPMs~\cite{ho2020denoising} are defined by two processes: The forward and the reverse. In the former, an input image $\mathbf{x}_0$ is gradually corrupted by adding Gaussian noise at each time step $t$, while the latter gradually removes it to generate a clean image. Formally, the forward process is defined by generating the time $t$ image $\mathbf{x}_t \sim \mathcal{N}\left(\sqrt{1-\beta_t} \mathbf{x}_{t-1}, \beta_t \mathbf{I}\right)$ iteratively from the original, clean image $\mathbf{x}_0$, where $\{\beta\}_{t=1: T}$ controls the variance of the noise added per step. The time $t$ image $\mathbf{x}_{t}$ can be sampled directly using $\mathbf{x}_0$ and the reparametrization trick,

\begin{equation}
    \mathbf{x}_t =\sqrt{\bar{\alpha}_t} \mathbf{x}_0+ \bm{\epsilon} \sqrt{1-\bar{\alpha}_t}, \bm{\epsilon} \sim \mathcal{N}(0,1),
    \label{eq:diffkernel}
\end{equation}

\noindent where $\alpha_t=1-\beta_t$ and $\overline{\alpha_t}=\prod_{s=1}^t \alpha_s$. %and $1 \leq t\leq T $. 
The reverse process also consists of Gaussian transitions whose mean and covariance are predicted by neural networks: $\mathbf{x}_{t-1} \sim \mathcal{N}\left(\mu_\theta\left(\mathbf{x}_t, t\right), \Sigma_\theta\left(\mathbf{x}_t, t\right) \right)$, where $\mathbf{x}_T \sim \mathcal{N}\left(\mathbf{0}, \mathbf{I}\right)$.
In practice, %variance is constant, $\sigma_t^2$, while 
a denoiser $\epsilon_\theta\left(\mathbf{x}_t, t\right)$ predicts the noise from Eq.~\eqref{eq:diffkernel} rather than predicting the mean $\mu_\theta\left(\mathbf{x}_t, t\right)$ directly, giving
$\mu_\theta\left(\mathbf{x}_t, t\right)=\frac{1}{\sqrt{1-\beta_t}}\left(\mathbf{x}_t-\frac{\beta_t}{\sqrt{1-\bar{\alpha}_t}} \epsilon_\theta\left(\mathbf{x}_t, t\right)\right)$.

\subsection{Diff-ICE: Diffusion-based Iterative Counterfactual Explanations}
We quantify image quality using a classifier $f$, which is trained to predict whether fetal ultrasound images are standard or non-standard (SP or NSP) planes.
As both ultrasound image quality~\cite{lin2022saw} and outcome prediction~\cite{pegios2023leveraging} benefit from combining images with segmentations, classifier $f$ consists of a segmentation model $s$ and a predictor $l$ trained sequentially. The classifier takes as inputs the image and the segmentation predictions, $f(\mathbf{x}) = l(s(\mathbf{x}), \mathbf{x})$. This adds explainability to the classifier, as the segmentations can be visualized as partial explanations.

Following~\cite{jeanneret2022diffusion,weng2024fast}, we corrupt the input $\mathbf{x}$ up to a limited noise level $\tau$, with $1 \leq \tau\leq T $, using Eq.\eqref{eq:diffkernel} to initialize a noisy version of the input and guide $\mathbf{x}^{c}_{\tau}$ towards the desired counterfactual class $y$ with guided diffusion~\cite{dhariwal2021diffusion}. To this end, we minimize a loss function $\mathcal{L}$ and shift the average sample with its gradient $g$,
\begin{equation}
\mathbf{x}^c_{t-1} \sim \mathcal{N}\left(\mu_\theta\left(\mathbf{x}^c_t, t\right) -  \Sigma_\theta\left(\mathbf{x}^c_t, t\right) g , \Sigma_\theta\left(\mathbf{x}^c_t, t\right) \right).
\label{eq:clsguidance}
\end{equation}

\noindent As the classifier $f$ is trained on clean images, we use the learned denoiser to get one-step denoised predictions and pass them through $f$ by solving Eq.\eqref{eq:diffkernel} for $\mathbf{x}_0$,
$\hat{\mathbf{x}}^c_{0|t}=\frac{\mathbf{x}^c_t-\sqrt{1-\alpha_t} \epsilon_\theta\left(\mathbf{x}^c_t, t\right)}{\sqrt{\alpha_t}}.
\label{eq:onestepdenoised}$
Computing the gradient w.r.t.~$\mathbf{x}^{c}_{t}$ as in~\cite{bansal2023universal, yu2023freedom}, i.e, $g = \nabla_{\mathbf{x}^{c}_{t}} \mathcal{L}$, necessitates backpropagating through the denoiser. As this is computationally expensive, we follow an efficient gradient estimation~\cite{he2023manifold,weng2024fast} which avoids excessive backpropagation and speeds up sampling.
Thus, for each $t$, we compute gradients w.r.t $\hat{\mathbf{x}}^{c}_{0|t}$, i.e,  $g = \nabla_{\hat{\mathbf{x}}^{c}_{0|t}} \mathcal{L}$ in Eq.~\eqref{eq:clsguidance} using counterfactual loss:

\begin{equation}
    \mathcal{L}(\mathbf{x}, \hat{\mathbf{x}}^{c}_{0|t}, y) = \lambda_c \mathcal{L}_c\left( f \left(\hat{\mathbf{x}}^{c}_{0|t}\right), y \right) + \lambda_p \mathcal{L}_p \left(\hat{\mathbf{x}}^{c}_{0|t}, \mathbf{x} \right),
    \label{eq:loss}
\end{equation}

\noindent where $\mathcal{L}_c$ is the classification loss which guides towards the desired label $y$, $\mathcal{L}_p$ is an l2-based perceptual loss which guides the process in terms of proximity, and $\lambda_c$ and $\lambda_p$ are hyperparameters which control the guidance strength. Typical applications focus on localized counterfactual changes and use a pixel-based l1-norm for $\mathcal{L}_p$~\cite{jeanneret2023adversarial, weng2024fast}  or this is added as an extra term on the noisy images~\cite{jeanneret2022diffusion}. Our loss function prioritizes broad changes while preserving anatomical fidelity. 

Yet, achieving global changes is challenging. Setting $\tau$ for a limited noise level preserves semantic information in one-step denoised predictions but allows guidance mostly in refinement stages~\cite{yu2023freedom}. Naively increasing the classifier guidance strength $\lambda_c$ may result in non-meaningful generations or adversarial examples~\cite{bansal2023universal}. Thus, we propose a \textbf{Diff}usion\textbf{-}based \textbf{I}terative \textbf{C}ounterfactual \textbf{E}xplanation approach (Diff-ICE) to enhance confidence and enable global alterations. Through $L$ iterations of the counterfactually guided reverse process, each using the previous output as input, we refine the output while maintaining fidelity constraints close to the original input $\mathbf{x}$ in Eq.~\eqref{eq:loss}. Our approach is summarized in Fig.~\ref{fig:method}.

\section{Experiments and Results }
\subsubsection{Data and base implementation.} We use two datasets extracted from a national fetal ultrasound screening database (ANONYMIZED). \texttt{GROWTH}, used to train the unconditional diffusion model and a segmentation model used in the guiding standard plane classifier $f$, consists of 4363 (2842/1521 for train/test) fetal ultrasound images from growth scans, including head, abdomen, femur, and cervix images. \texttt{HEAD} is used to train and test the full guiding classifier $f$, and consists of fetal \emph{head} ultrasound images which include 240 high-quality standard planes (SP) and 1339 low-quality non-standard planes (NSP). As the guiding \emph{standard plane classifier} $f$ we choose a robust and interpretable architecture that combines a DTU-Net~\cite{lin2023dtu} segmentation model $s$ with a SonoNet-16~\cite{baumgartner2017sononet} classifier $l$ following~\cite{pegios2023leveraging}. Robustness is important to ensure high-quality counterfactuals, and interpretability makes the counterfactuals easier to monitor both for technical developers and clinicians at different levels of experience. The segmentation model $s$ is developed on \texttt{GROWTH}. Thus, we train and evaluate the classifier's predictor $l$ sequentially, keeping the weights of $g$ fixed on a split of 121/26/93 SP and 712/204/423 NSP images for train/validation/test with non-overlapping patients, resulting in $78\%$ balanced test accuracy. An unconditional \emph{DDPM}~\cite{ho2020denoising} is also trained on \texttt{GROWTH} using 1000 diffusion steps, following the architecture of ~\cite{jeanneret2022diffusion}, 
training for 300K iterations with batch size 16, learning rate $10^{-4}$, weight decay of 0.05, and no dropout. For all models, images are resized to $224 \times 288$, embedded text and calipers are removed~\cite{mikolaj2023removing}, and pixel intensity is normalized to $[-1,1]$.  To generate \emph{counterfactual visual explanations}, we empirically set $L=5$, $\tau = 120$ of 400 re-spaced time steps, $\lambda_{p}=30$ and search for $\lambda_{c} \in \{40,60,80\}$. The perceptual loss uses a ResNet-50 trained on RadImageNet~\cite{mei2022radimagenet}.

\begin{table}[t]
\caption{Comparison of Diff-ICE with baseline diffusion-based approaches.} 
\label{tab:main}
\centering
\scriptsize
\resizebox{\linewidth}{!}{
\begin{tabular}{c|ccc|cccc|ccc}
\toprule

\multicolumn{1}{c|}{} & \multicolumn{3}{|c|}{Realism} & \multicolumn{1}{c}{} & \multicolumn{2}{c}{Validity} &  \multicolumn{1}{c}{} & \multicolumn{3}{|c}{Efficiency} \\\midrule

Method & FID $\downarrow$ & FSD $\downarrow$ & SonoSim $\uparrow$ & $\text{MQD}$ $\uparrow$ & BKL $\downarrow$ & MAD $\uparrow$ & FR $\uparrow$ & Batch Time (s) & Total Time (h) & GPU M (GB)\\
%&  &  &  &  &  & & &  Time (s)  &  Time (h) & (GB)\\
\midrule\midrule
\dime & 41.5 & 0.396 & 0.854 & 0.291 & 0.391 & 0.231 & 0.966 & 3151.9 {$\pm$ 730.4} & 37.65 & \textbf{9.6} \\
\midrule
\difficeonenoisy & 39.9 & 0.403 & 0.855 & 0.301 & 0.413 & 0.208 & 0.936 & 231.4 {$\pm$ 56.8} & 2.77 & 33.7 \\
\midrule
\difficeone & \textbf{39.0} & \textbf{0.355} & \textbf{0.856} &  0.253 & 0.387 & 0.234   & \textbf{0.982} & \textbf{115.6 {$\pm$ 34.2}} & \textbf{1.38} & \textbf{9.6} \\
\midrule\midrule
\diffice & 42.4 & 0.435 & 0.790 & \textbf{0.371} & \textbf{0.336} & \textbf{0.284} & \textbf{0.982} & 448.6 {$\pm$ 22.9} & 5.27 & \textbf{9.6} \\
\bottomrule
\end{tabular}}
\label{tab:comparison}
\end{table}

\begin{table}[t]
\caption{Intermediate results for each iteration of \diffice.} 
\label{tab:steps}
\centering
\scriptsize
\resizebox{\linewidth}{!}{
\begin{tabular}{c|ccc|ccc|c|cc}
\toprule
\multicolumn{1}{c|}{} & \multicolumn{3}{|c|}{Realism} & \multicolumn{3}{c|}{Validity} & \multicolumn{1}{c|}{Re-identification} &\multicolumn{2}{|c}{Efficiency (Run Time)} \\\midrule

Iteration & FID $\downarrow$ & FSD $\downarrow$ & SonoSim $\uparrow$ & $\text{MQD}$ $\uparrow$ & BKL $\downarrow$ & MAD $\uparrow$ & Accuracy (\%) $\uparrow$ & Batch (s) & Iteration (h)\\
\midrule\midrule
1 &  \textbf{38.96} & \textbf{0.355} & \textbf{0.856} &  0.253 & 0.387 & 0.234  & \textbf{99.76} & 115.6 {$\pm$ 34.2} & 1.38 \\
\midrule
2 & 41.26 & 0.420 & 0.830 & 0.317 & 0.370 & 0.250 & 98.80 & 88.6 {$\pm$ 22.3} & 1.06  \\
\midrule
3 & 42.25 & 0.464 & 0.818 &  0.314 & 0.362 & 0.258    & 97.36 &85.8 {$\pm$ 21.9} & 1.03 \\
\midrule
4 & 43.17 & 0.454 & 0.807 & 0.355 & 0.357 & 0.262  & 94.48 &77.9 {$\pm$ 18.7} & 0.96 \\
\midrule
5 & 42.42 & 0.435 & 0.790 & \textbf{0.371} & \textbf{0.336} & \textbf{0.284} & 91.83 &78.9 {$\pm$ 17.4} & 0.95 \\
\bottomrule
\end{tabular}}
\label{tab:iterations}
\end{table}

\subsubsection{Baselines.}
1) \dime~\cite{jeanneret2022diffusion} employs an expensive nested loop of reverse guided processes per step $t$ to obtain clean images and applies a scaling to estimate gradients w.r.t.~noisy images. 2) A single iteration of \diffice, denoted as  \difficeone. 3) Inspired by~\cite{bansal2023universal} we implement \difficeonenoisy taking the gradient w.r.t noisy images. For fair comparison, we use the same loss (Eq.~\eqref{eq:loss}) and hyperparameters.

\begin{figure}[t]
	\centering
    \subfloat{\includegraphics[scale=0.19]{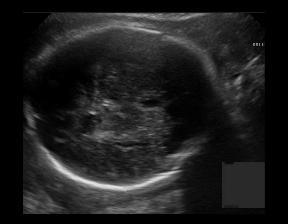}} \hspace{0.1px} 
	\subfloat{\includegraphics[scale=0.19]{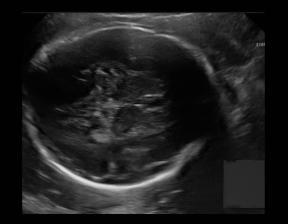}} \hspace{0.1px}
	\subfloat{\includegraphics[scale=0.19]{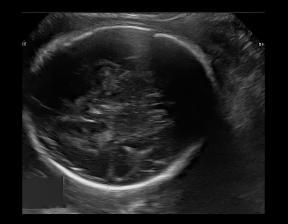}} \hspace{0.1px}
    \subfloat{\includegraphics[scale=0.19]{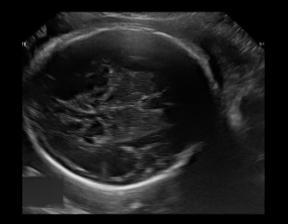}} \hspace{0.1px}
    \subfloat{\includegraphics[scale=0.19]{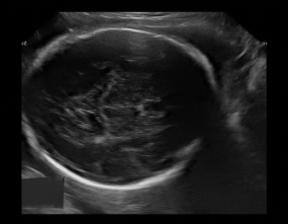}}
    \hspace{0.1px}
    \subfloat{\includegraphics[scale=0.19]{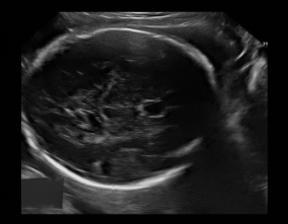}}
    
    \vspace{-1.1em}

 	\subfloat{\includegraphics[scale=0.19]{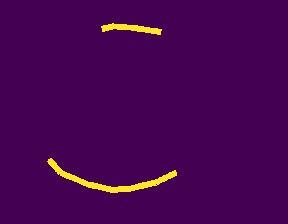}} \hspace{0.1px} 
	\subfloat{\includegraphics[scale=0.19]{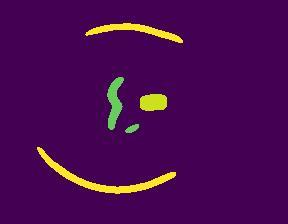}} \hspace{0.1px}
	\subfloat{\includegraphics[scale=0.19]{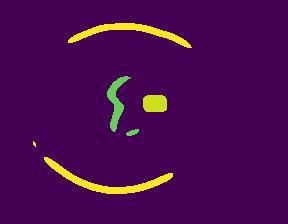}} \hspace{0.1px}
    \subfloat{\includegraphics[scale=0.19]{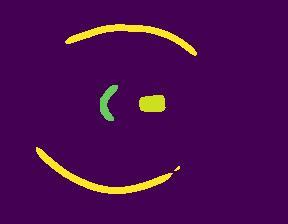}} \hspace{0.1px}
    \subfloat{\includegraphics[scale=0.19]{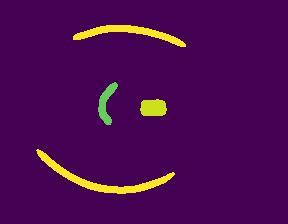}} \hspace{0.1px}
    \subfloat{\includegraphics[scale=0.19]{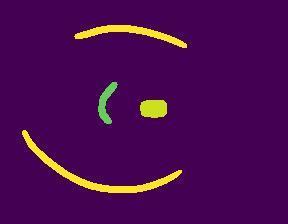}}

    \vspace{0.2em}
    \hrule
    \vspace{-0.9em}

    \subfloat{\includegraphics[scale=0.19]{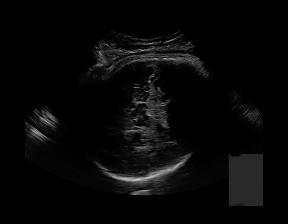}} \hspace{0.1px} 
	\subfloat{\includegraphics[scale=0.19]{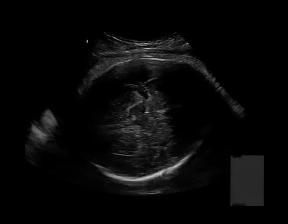}} \hspace{0.1px}
	\subfloat{\includegraphics[scale=0.19]{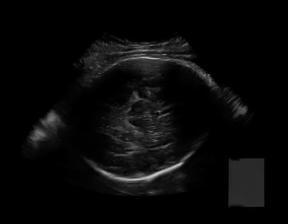}} \hspace{0.1px}
    \subfloat{\includegraphics[scale=0.19]{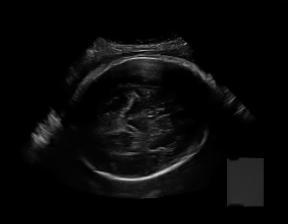}} \hspace{0.1px}
    \subfloat{\includegraphics[scale=0.19]{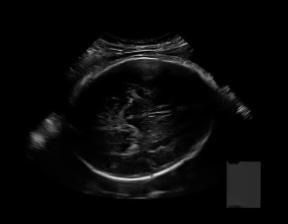}}
    \hspace{0.1px}
    \subfloat{\includegraphics[scale=0.19]{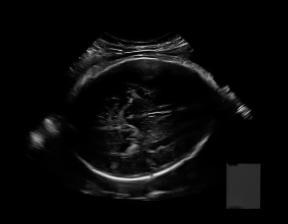}}
    
    \vspace{-1.1em}

 	\subfloat{\includegraphics[scale=0.19]{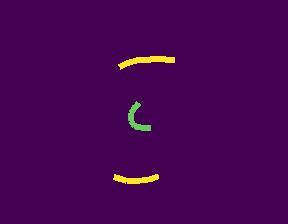}} \hspace{0.1px} 
	\subfloat{\includegraphics[scale=0.19]{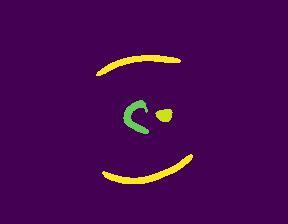}} \hspace{0.1px}
	\subfloat{\includegraphics[scale=0.19]{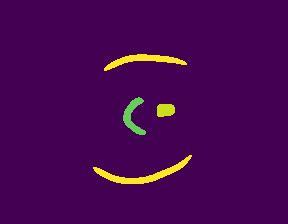}} \hspace{0.1px}
    \subfloat{\includegraphics[scale=0.19]{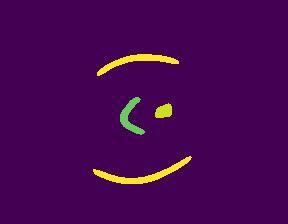}} \hspace{0.1px}
    \subfloat{\includegraphics[scale=0.19]{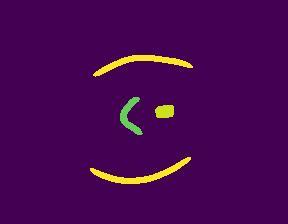}} \hspace{0.1px}
    \subfloat{\includegraphics[scale=0.19]{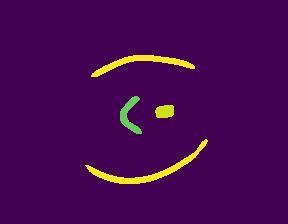}}

    \vspace{0.2em}
    \hrule
    \vspace{-0.9em}
    
    \subfloat{\includegraphics[scale=0.19]{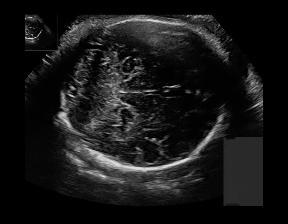}} \hspace{0.1px} 
	\subfloat{\includegraphics[scale=0.19]{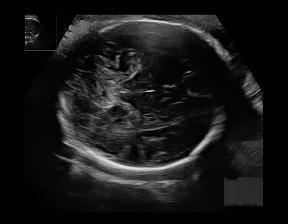}} \hspace{0.1px}
	\subfloat{\includegraphics[scale=0.19]{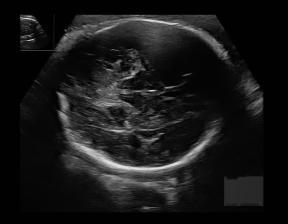}} \hspace{0.1px}
    \subfloat{\includegraphics[scale=0.19]{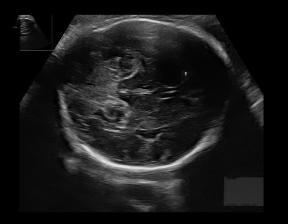}} \hspace{0.1px}
    \subfloat{\includegraphics[scale=0.19]{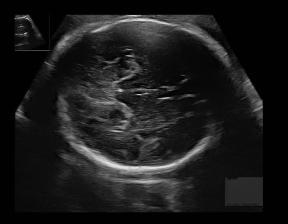}}
    \hspace{0.1px}
    \subfloat{\includegraphics[scale=0.19]{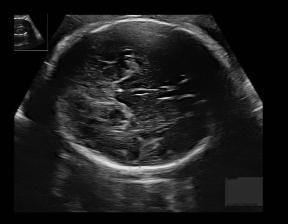}}
    
    \vspace{-1.1em}
    \setcounter{subfigure}{0}
 	\subfloat[ \scriptsize NSP Input]{\includegraphics[scale=0.19]{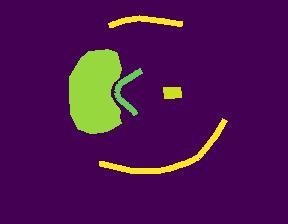}} \hspace{0.1px} 
	\subfloat[\scriptsize Iter 1]{\includegraphics[scale=0.19]{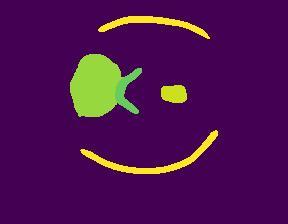}} \hspace{0.1px}
	\subfloat[\scriptsize Iter 2]{\includegraphics[scale=0.19]{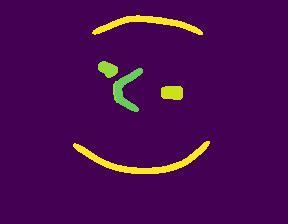}} \hspace{0.1px}
    \subfloat[\scriptsize Iter 3]{\includegraphics[scale=0.19]{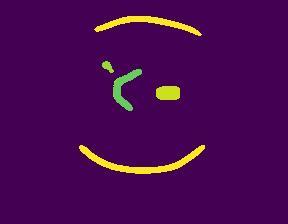}} \hspace{0.1px}
    \subfloat[\scriptsize Iter 4]{\includegraphics[scale=0.19]{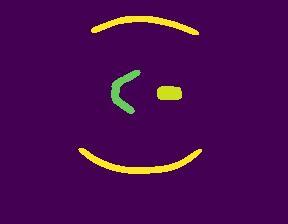}} \hspace{0.1px}
    \subfloat[\scriptsize Iter 5]{\includegraphics[scale=0.19]{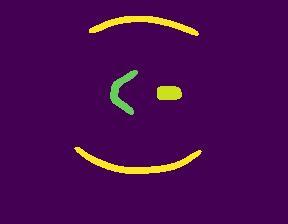}}

\caption{Iterations of \diffice from the low-quality NSP to a higher-quality SP, visualized with predicted segmentations for interpretability (expert mask annotation for NSP input). Diff-ICE synthesizes planes that (correctly) contain \textcolor{thalamus}{thalamus (TH)}, and \textcolor{cavum}{cavum septi pellucidi (CSP)} from NSP planes. It is also capable of creating planes that (correctly) remove \textcolor{fossa}{fossa posterior (FP)}. In addition to these explicit changes, Diff-ICE achieves broad changes and refinements.
}
 \label{fig:examples}
\end{figure}

\subsubsection{Evaluation metrics.} 
Following existing evaluation protocols~\cite{jeanneret2022diffusion,jeanneret2023adversarial,weng2024fast,lee2020generating}, we evaluate \emph{realism} using Fréchet Inception Distance (FID) and Fréchet SonoNet Distance~\cite{lee2020generating} (FSD) between the original NSP images and their valid SP counterfactuals. We further introduce SonoSim, which computes the cosine similarity using SonoNet-64~\cite{baumgartner2017sononet} features, similar to those used in FSD. To measure \emph{validity} of generated images, we use standard metrics from computer vision such as Flip Ratio (FR), i.e., the frequency of counterfactuals classified as SP, Mean Absolute Difference~\cite{weng2024fast} (MAD) of confidence prediction between original NSP and counterfactual SP, and Bounded remapping of KL divergence~\cite{jeanneret2022diffusion} (BKL), which measures similarity between the counterfactual's prediction by $f$, and the SP one-hot counterfactual label. %, with lower values indicating higher similarity. 
The validity metrics %BKL, MAD, and FR 
are computed for those NSP test images that are classified as NSP by $f$. In addition to verifying that counterfactuals indeed move towards the SP class according to the guiding classifier $f$, we develop a progressive concept bottleneck model~\cite{lin2022saw} (PCBM)  as an \emph{oracle} using \texttt{GROWTH}, and use its confidence to measure overall Quality Scores ($\text{QS}_{\text{O}}$) for both input $\mathbf{x}$ and counterfactual $\mathbf{x}^{c}$. To simulate a realistic evaluation scenario and ensure reliable oracle predictions for our analysis, we include cases with confident predictions of the original NSP input, i.e., original inputs $\mathbf{x}$ classified as NSP, $\text{QS}_{\text{O}}(\mathbf{x}) < 0.5 $.
As an \emph{oracle validity} metric, we introduce Mean overall Quality Difference defined as
$\text{MQD} = \frac{1}{N} \sum_{i=1}^{N} I(\text{QS}_{\text{O}}(\mathbf{x}_{i}) < 0.5) \cdot \text{QD}_{\text{O}}(\mathbf{x}^{c}_{i}, \mathbf{x}_{i})$ where $I$ is the indicator function, $N=423$ NSP test images and $\text{QD}_{\text{O}}(\mathbf{x}^{c}, \mathbf{x}) = \text{QS}_{\text{O}}(\mathbf{x}^{c}) - \text{QS}_{\text{O}}(\mathbf{x})$. To evaluate \emph{efficiency}, we compute batch time in seconds, total time in hours, and GPU memory, using a batch size of 10 on an NVIDIA RTX A6000.

\subsubsection{Results.} 
\begin{figure}[t]
	%\centering
	\subfloat{\includegraphics[width = 0.24\linewidth, height =0.22\linewidth]{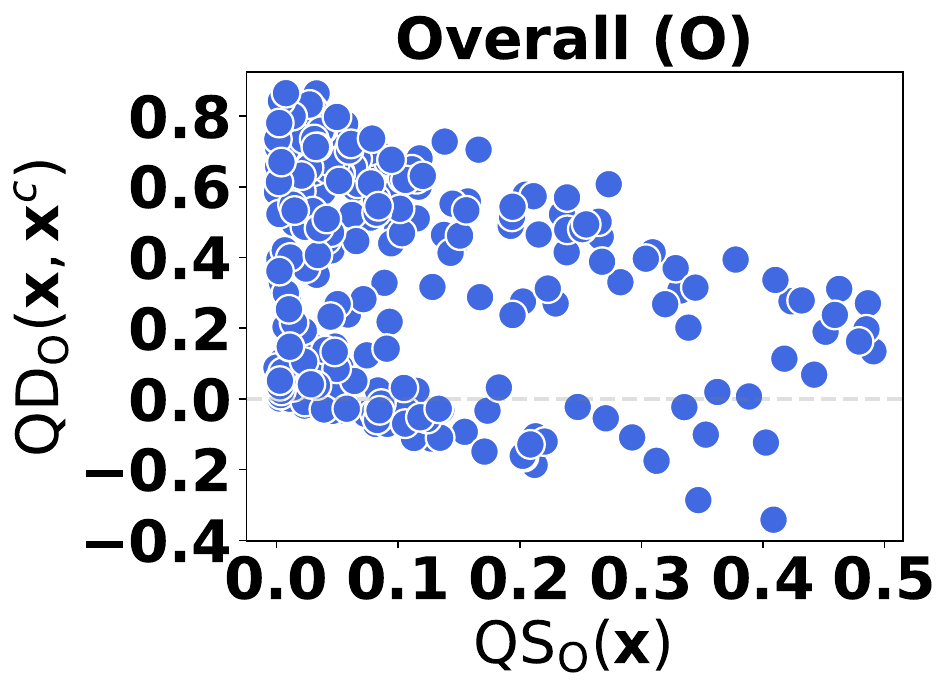}} \hspace{0.1px} 
	\subfloat{\includegraphics[width = 0.24\linewidth, height =0.22\linewidth]{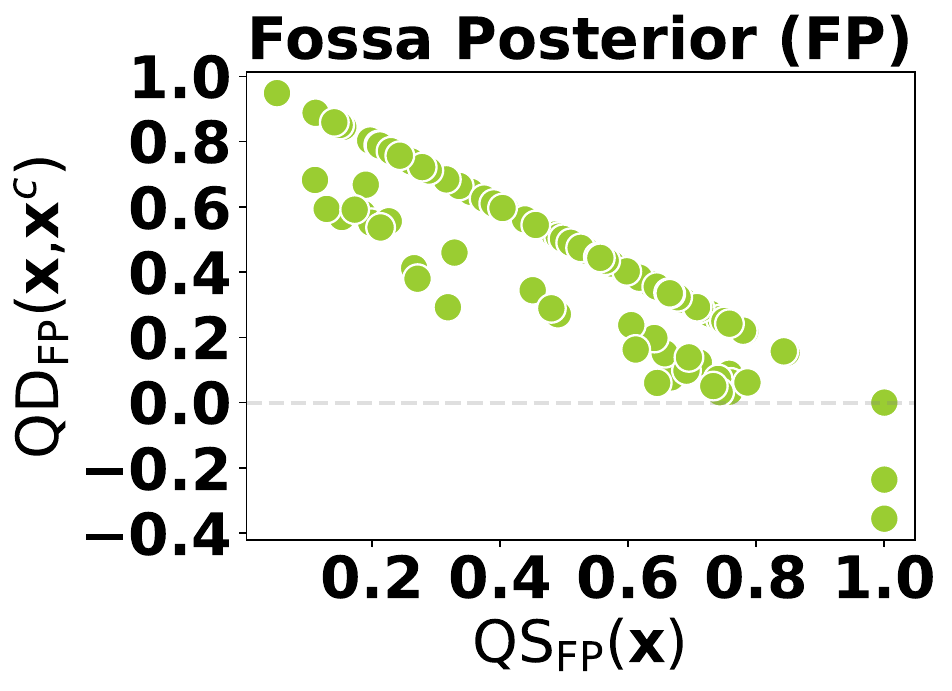}}\hspace{0.1px}
	\subfloat{\includegraphics[width = 0.24\linewidth, height =0.22\linewidth]{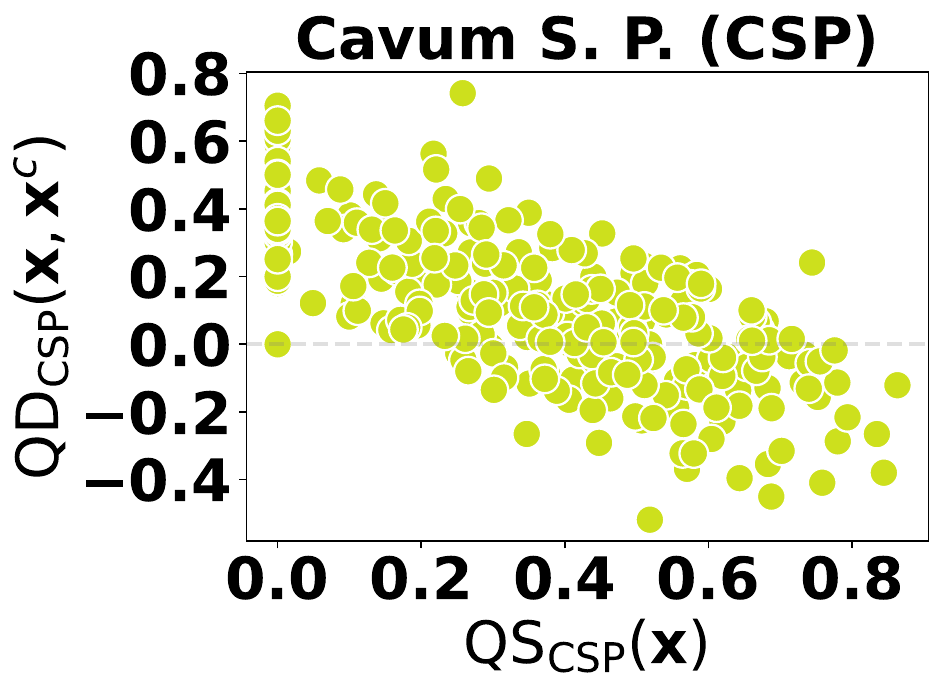}} \hspace{0.1px} 
	\subfloat{\includegraphics[width = 0.24\linewidth, height =0.22\linewidth]{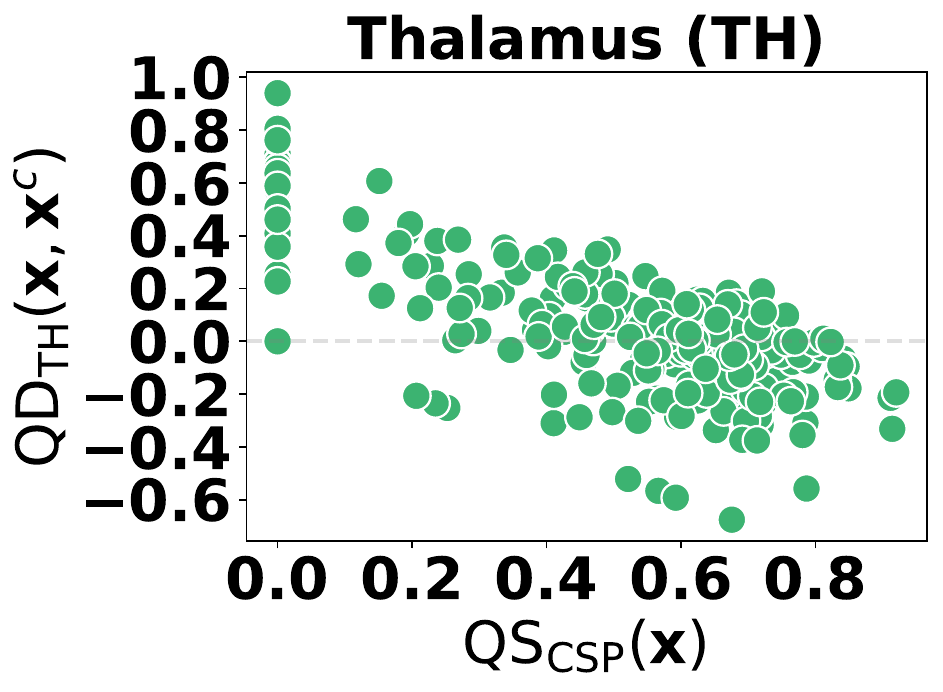}}

	\caption{Quality Difference (QD) as a function of NSP input's Quality Score (QS).}
 \label{fig:qd}
\end{figure}

We generated counterfactuals for all 423 NSP test images from \texttt{HEAD}. Tables~\ref{tab:comparison} and~\ref{tab:iterations} list results for all methods and each intermediate iteration of \diffice, respectively.
Table~\ref{tab:iterations} illustrates the expected trade-off, with realism decreasing as validity increases with more iterations. To assess \emph{identity preservation}~\cite{gabler2023fetal}, we use SonoSim and report rank-3 (re-identification) accuracy on 417 valid counterfactuals from 308 unique fetuses, verifying whether the fetal identity of SP counterfactual image is among the top 3 most similar input NSP images. We set $L=5$ due to the small differences between the last iterations, as shown in Fig.~\ref{fig:examples}, which illustrate paths from low-quality NSP to higher-quality SP.
In addition to image-level SP confidence scores, the oracle PCBM~\cite{lin2022saw} also predicts individual quality scores ($\text{QS}_{\text{FP}}$, $\text{QS}_{\text{CSP}}$ and $\text{QS}_{\text{TH}}$ for \textcolor{fossa}{fossa posterior (FP)}, \textcolor{cavum}{cavum septi pellucidi (CSP)}, and \textcolor{thalamus}{thalamus (TH)}). As a fine-grained assessment of \diffice, we show their quality score differences between counterfactual and NSP input ($\text{QD}_{\text{FP}}$, $\text{QD}_{\text{CSP}}$, $\text{QD}_{\text{TH}}$) as a function of input’s quality in Fig.~\ref{fig:qd}.

\begin{figure}[t]
	\centering
    
    \subfloat{\includegraphics[scale=0.22]{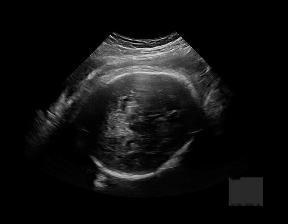}} \hspace{0.1px} 
	\subfloat{\includegraphics[scale=0.22]{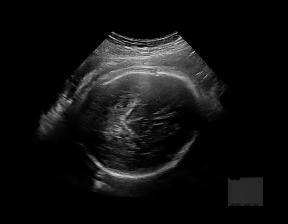}} \hspace{0.1px}
     \subfloat{\includegraphics[scale=0.22]{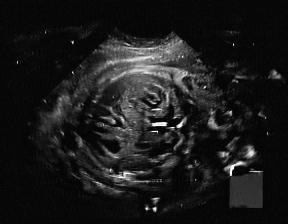}} \hspace{0.1px}
    \subfloat{\includegraphics[scale=0.22]{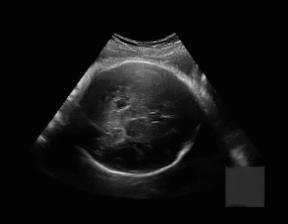}} \hspace{0.1px}
    \subfloat{\includegraphics[scale=0.22]{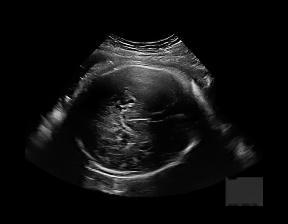}}

    \vspace{-1.1em}
    
	\subfloat{\includegraphics[scale=0.22]{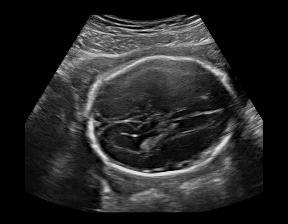}} \hspace{0.1px} 
	\subfloat{\includegraphics[scale=0.22]{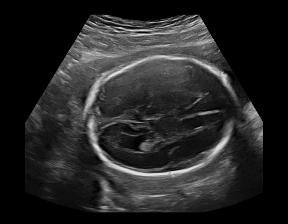}} \hspace{0.1px}
     \subfloat{\includegraphics[scale=0.22]{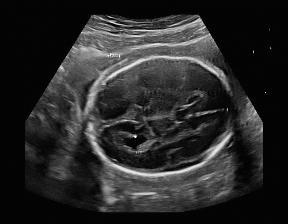}} \hspace{0.1px}
    \subfloat{\includegraphics[scale=0.22]{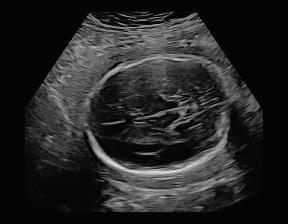}} \hspace{0.1px}
    \subfloat{\includegraphics[scale=0.22]{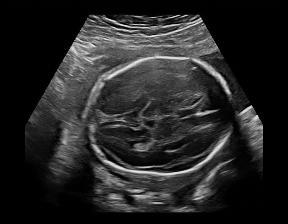}}
    
    \vspace{-1.1em}

    \setcounter{subfigure}{0}

     \subfloat[\scriptsize NSP Input]{\includegraphics[scale=0.22]{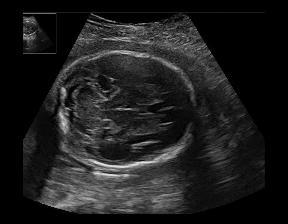}} \hspace{1px} 
	\subfloat[\scriptsize $\tau = 80$]{\includegraphics[scale=0.22]{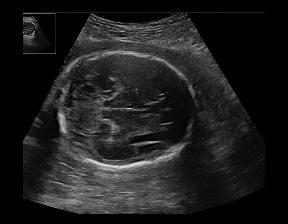}} \hspace{0.1px}
  \subfloat[\scriptsize $\text{\medspace\medspace\medspace}\tau = 120$, \\$\text{\medspace\medspace\medspace\medspace\medspace\medspace\medspace\medspace\medspace}\lambda_c = 400 $]{\includegraphics[scale=0.22]{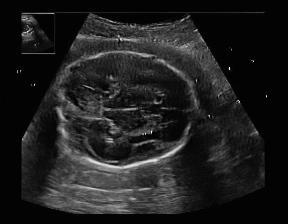}} \hspace{0.1px} 
 \subfloat[\scriptsize $\tau = 200$]{\includegraphics[scale=0.22]{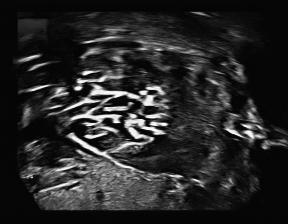}} \hspace{0.1px} 
 \subfloat[\scriptsize Diff-ICE]{\includegraphics[scale=0.22]{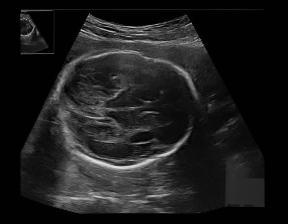}}

	\caption{\textbf{Iterative counterfactuals.} \difficeone is illustrated in columns (b), (c), and (d). Increasing $\lambda_c$ (c) can lead to adversarial noises. Low values for $\tau$ (b) lead to insignificant changes, while high values (d) can completely change anatomy.}
 \label{fig:ablation1}
\end{figure}

\subsubsection{On the need for iterative counterfactuals.}
We assessed the impact of parameters $\tau$ (initialize noise level) and $\lambda_c$ (guidance strength), in our single iteration baseline \difficeone, relative to our iterative method \diffice. 
As illustrated in Fig.~\ref{fig:ablation1}, we observed that trying to achieve global changes by naively increasing guidance strength leads to adversarial examples, while high levels of initial noise can result in changes in the anatomy or collapse to adversarial examples. In contrast, \diffice effectively improves quality while maintaining image fidelity. 

\subsubsection{Qualitative validation.} A fetal medicine consultant with 10 years of experience in obstetric imaging was asked to select the best quality image from pairs of real NSP images and associated \diffice counterfactuals. Image pairs with $\text{QD}_{\text{O}}>0$ were sampled
uniformly and displayed in randomized positions and order. The expert selected \diffice counterfactuals in 41 out of 50 image pairs, demonstrating the ability of \diffice to indeed enhance the input's quality. The expert also stated qualitatively that differences were global, including improved presentation of bone outlines, \textcolor{thalamus}{TH}, and \textcolor{cavum}{CSP}, and improved overall image quality.

\section{Discussion and Limitations}
Our results show that our iterative approach produces counterfactuals with higher confidence and broader changes than single-step baselines, while largely preserving image fidelity and fetal identity. Moreover, we maintain computational feasibility through the adopted efficient gradient estimation scheme. 
Fig.~\ref{fig:qd} shows that improvement is largest when the initial quality is poor, which is intuitive, as changes to high-quality inputs close to the SP should be minimal. For all structures, however, the quality is generally improved. \diffice is especially effective at removing \textcolor{fossa}{FP}, whereas in cases where \textcolor{thalamus}{TH} and \textcolor{cavum}{CSP} are fairly well-represented, it balances overall quality with the quality of individual structures.

While our results are promising in terms of image fidelity and fetal ultrasound-specific metrics, the absence of datasets containing both NSP and SP for the same fetus limited our ability to perform direct comparisons, assess hallucinations, and evaluate the impact on biometric measurements -- all crucial for clinical validation. Nevertheless, the ability to visualize plausible paths toward SP offers strong potential for personalized education, supporting the training of sonographers and enabling non-experts to annotate datasets~\cite{chiquier2025teaching}. Future work could integrate fetus-specific priors from video scans when available or directly adapt counterfactual video generation approaches~\cite{spyrou2025causally}, and focus on clinical evaluations towards supporting SP acquisition for downstream diagnosis and monitoring. 

\subsubsection{Acknowledgements.} 
This work was supported by the Pioneer Centre for AI (DNRF grant nr P1), the DIREC project EXPLAIN-ME (9142-00001B), and the Novo Nordisk Foundation through the Center for Basic Machine Learning Research in Life Science (NNF20OC0062606). P.Pegios would like to thank Thanos Delatolas for insightful discussions.

\bibliographystyle{splncs04}
\bibliography{ref.bib}

\end{document}